\documentclass[cameraready]{Interspeech}

\title{Exploring Pre-training Benefits on Phoneme Addition \\
through Fine-tuning in Speech Synthesis}

\author[affiliation={1,2}, orcid=0009-0002-8344-0783, correspondingauthor]{Masato}{Murata}
\author[affiliation={1}, orcid=0000-0002-5796-4535]{Koichi}{Miyazaki}
\author[affiliation={1}, orcid=0000-0002-8347-5604]{Tomoki}{Koriyama}
\author[affiliation={2}, orcid=0000-0001-8146-1279]{Tomoki}{Toda}

\address{
    $^1$ CyberAgent, Japan 
    $^2$ Nagoya University, Japan 
}

\email{murata\_masato@cyberagent.co.jp, miyazaki\_koichi\_xa@cyberagent.co.jp, koriyama\_tomoki@cyberagent.co.jp, tomoki@icts.nagoya-u.ac.jp}

\keywords{Speech synthesis, text-to-speech, phoneme addition, fine-tuning,  transfer learning, cross-lingual TTS}

\usepackage{comment}
\usepackage{cite}
\usepackage{multirow}

\begin{document}

\maketitle

\begin{abstract}

Transfer learning is widely used for low-resource text-to-speech. When the target corpus contains phonemes unseen in pre-training, the model must expand its phoneme inventory during fine-tuning; we call the process ``phoneme addition.'' However, it remains unclear whether the pre-trained ability to generate seen phonemes contributes to this process.
This study investigates phoneme addition in two settings: (1) a simulation setup using LLM-generated phoneme-controlled corpora that enables investigation without considering confounding factors, and (2) a real-speech cross-lingual transfer setup (English to Japanese) to validate whether the findings hold in practice.
Experiments in both settings showed that while fine-tuning achieved higher naturalness than training from scratch, it required as much or more data to achieve comparable PER for new phonemes. These results indicate that pre-training mainly contributes to naturalness improvement, but offers limited benefit for phoneme addition.

\end{abstract}

\section{Introduction}
Recent text-to-speech (TTS) systems have achieved human-level naturalness in high‑resource scenarios.
However, developing low-resource language TTS remains challenging.
To solve this data limitation issue, transfer learning is a standard approach, where a model is pre-trained on a high-resource language and subsequently fine-tuned on the target language.
In practice, the target dataset often contains new phonemes unseen in pre-training, requiring the model to expand its phoneme inventory to add the phonemes during fine-tuning.
This research focuses on this process, referred to as ``phoneme addition.''

Phoneme addition is especially common in cross-lingual adaptation, where the source language phoneme inventory often lacks specific phonemes required for target languages.
For example, when building Spanish TTS from an English pre-trained model, the model requires adding alveolar trills (rr), and for the Xhosa language requires adding click consonants (c, q, and x).
In this study, we refer to such new phonemes to be acquired during fine-tuning as ``target phonemes,'' and define the pre-trained ability to generate seen phonemes as ``pre-trained phoneme knowledge.''
These transfer learning approaches for low-resource language TTS rely on the assumption that the linguistic and acoustic knowledge acquired during pre-training contributes to fine-tuning. 
Previous studies on low-resource language TTS have addressed these unseen target phonemes by phoneme mapping~\cite{chen2019end, do2022text, huang2022few, do2023strategies, byambadorj2021text} or random initialization~\cite{amalas2024multilingual_randominit} before fine-tuning.
However, these studies reported only improvements in overall naturalness and intelligibility, leaving it unclear whether pre-trained phoneme knowledge really contributes to the phoneme addition process itself. 

\begin{figure}[t]
   \centering
    \begin{minipage}[b]{0.9\linewidth}
      \centering
      \centerline{\includegraphics[width=\hsize]{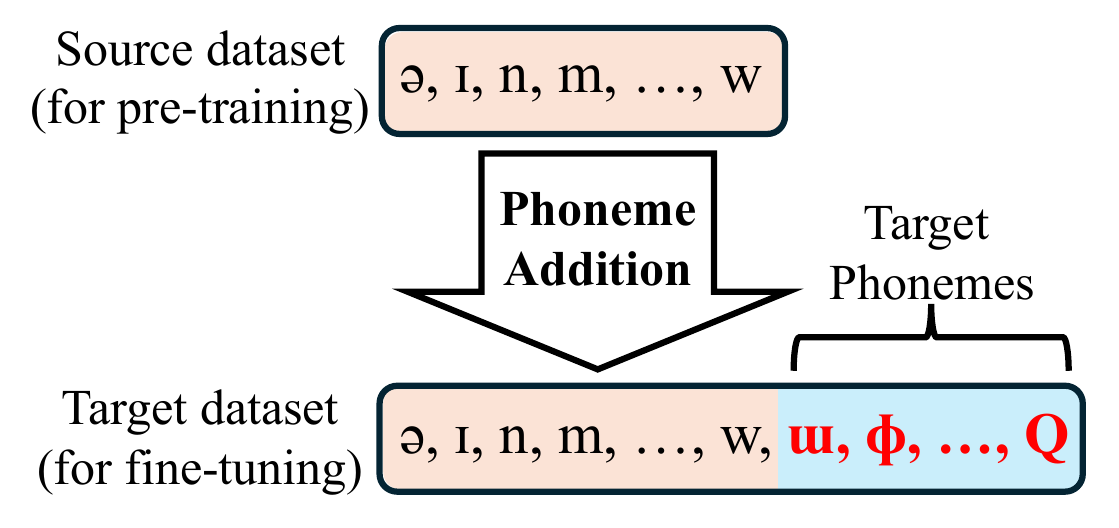}}
    \end{minipage}
        \vspace{-5pt} %
\caption{Overview of the phoneme addition process in transfer learning. The pre-trained model is trained on a source dataset. During fine-tuning, the phoneme inventory is expanded by adding target phonemes (red).}
    \vspace{-10pt} %
    \label{fig:overview}
  \end{figure}

To verify the benefits of pre-trained phoneme knowledge, we investigate this phoneme addition process by comparing fine-tuning and scratch training on the same target data.
This investigation is expected to lead to more efficient low-resource TTS development.
In this study, we conduct the investigation under two experimental settings: phoneme-controlled simulation and real-speech cross-lingual transfer.
First, we design a simulated setting using two phoneme-controlled corpora generated with a large language model (LLM): a ``Limited'' corpus excluding the specific target phonemes for pre-training, and a ``Full'' corpus containing all phonemes (including target phonemes) for fine-tuning.
These datasets enable us to simulate the phoneme addition process across various conditions, such as training dataset sizes and target phoneme types.
Moreover, they share the same language and target speaker, ensuring isolation from confounding factors such as language mismatch, speaker variation, and data domain differences.
Furthermore, we validate our simulation findings in a cross-lingual transfer learning setting (from English to Japanese) using real speech datasets to ensure practical applicability. 

Our experimental results reveal that TTS models can successfully acquire unseen phonemes through fine-tuning.
However, from the perspective of phoneme addition, fine-tuning requires similar or more data than training from scratch to achieve the same phoneme error rate (PER) for the target phonemes, regardless of the target phoneme type.
We observed this consistent result across both the phoneme-controlled simulation and the real-speech cross-lingual setting. 
Although fine-tuning starts from pre-trained models that have already acquired the ability to generate seen phonemes, the results indicate that the contribution of pre-trained phoneme knowledge is limited for phoneme addition.

The main contributions of this paper are as follows:
\begin{itemize}
\item We introduce LLM-generated phoneme-controlled corpora for a simulated setting that isolates the phoneme addition process from confounding factors, enabling controlled investigation across dataset sizes and target phoneme types.
\item Across both simulated and real-speech cross-lingual transfer settings, we found that pre-training mainly contributes to speech naturalness improvement, but offers limited benefit for phoneme addition compared to scratch training.
\end{itemize}

\section{Phoneme-Controlled Corpus Generation Using LLM}
\label{sec:method}
To investigate how pre-trained phoneme knowledge contributes to unseen phoneme addition through fine-tuning, we design a phoneme-controlled experimental setup that simulates the phoneme addition process.
These simulation settings enable us to investigate the phoneme addition process without considering confounding factors such as language mismatch and dataset domain differences.
We create two types of synthetic corpora by using an LLM: (1) a \textbf{Limited Corpus} that excludes specific target phonemes intended to be added during fine-tuning, and (2) a \textbf{Full Corpus} that contains all phonemes, including the target phonemes. 
This research uses Claude Opus 4.6~\cite{anthropic2026claude_opus_46} to generate the corpora.
We also use the statistics of the real speech dataset (VCTK~\cite{vctk}) to ensure the synthetic corpora closely simulate the real dataset.
To control the phoneme distribution, all generated sentences are verified by converting to IPA using espeak-ng\footnote{\url{https://github.com/espeak-ng/espeak-ng}} and checking the presence of target phonemes. 
The rule-based filtering successfully ensures that both corpora meet the phoneme constraints: the Limited corpus excludes all target phonemes, while the Full corpus contains the complete set of 40 English phonemes.
In this paper, we refer to these LLM-generated texts as ``corpora,'' and the synthesized speech--phoneme pairs as ``datasets.''

\subsection{Limited Corpus for Pre-training}
\label{subsection:limited_corpus_generation}
To simulate a pre-training scenario where specific target phonemes remain unseen, we create the Limited Corpus through the following process. We first extract an allowed word list (over 23,000 words) from the CMU Pronouncing Dictionary~\cite{cmudict} by selecting only words that do not contain any of the target phonemes. 
We then prompt the LLM to generate natural English sentences of 3--15 words using only words from this allowed list. 
Each generated sentence is verified by converting to IPA using espeak-ng and confirming the absence of target phonemes. 
Sentences failing this verification are discarded.
From this verified sentence pool, we sample 1,000 sentences to match the VCTK statistics such as the average number of phonemes per utterance and the word-count distribution.
Finally, we use a prepared TTS model to synthesize speech from the resultant IPA sentences, obtaining paired speech--IPA transcription data for pre-training.

\subsection{Full Corpus for Fine-tuning}
\label{subsection:full_corpus_generation}
We create the Full Corpus for the fine-tuning dataset, designed to enable the fine-tuned model to acquire the ability to generate new target phonemes.
We first generate a large sentence pool by prompting the LLM to generate natural English sentences with the word count distribution matched to the VCTK dataset.
After deduplication filtering, we verify that every sentence contains at least one target phoneme by converting each sentence to IPA using espeak-ng.
As a result, we obtain approximately 30,000 valid sentences.   
Following the Limited Corpus generation, we sample 2,000 sentences from this verified sentence pool to ensure that the synthetic corpus closely simulates the statistics of the real speech corpus. 
We then synthesize speech from the selected IPA sentences using the same TTS model as in the Limited Corpus generation, resulting in paired speech--IPA transcription data for fine-tuning.

Table~\ref{table:stats_corpora} shows statistics of the created phoneme-controlled corpora, Full and Limited, and the reference VCTK dataset, with plosive consonants and front vowels as target phonemes.
All statistics are computed over 1,000 randomly sampled sentences.
The resultant simulation corpora closely match the statistics of the real speech dataset.

\begin{table}[t!]
    \centering
    \caption{Statistics of the created phoneme-controlled corpora (Full and Limited) and VCTK dataset, computed over 1,000 sampled sentences. These synthetic corpora simulate the statistics of the VCTK dataset.}
    \label{table:stats_corpora}
    \scalebox{0.71}{
        \begin{tabular}{l l c c c c c} 
            \toprule
            & & \multicolumn{2}{c}{Num. Unique} & \multicolumn{3}{c}{Num. Avg (/Utt)} \\
            \cmidrule(lr){3-4} \cmidrule(lr){5-7}
            \multirow{2}{*}{Dataset} & \multirow{2}{*}{Target Type} & \multirow{2}{*}{Phonemes} & \multirow{2}{*}{Words} & \multirow{2}{*}{Phonemes} & Target & \multirow{2}{*}{Words} \\
            & & &  &  & Phoneme & \\
            \midrule
            \multirow{2}{*}{VCTK} 
            & Plosives     & 40 & 1,743 & 26.1 & 4.83 & 7.4 \\
            & Front vowels & 40 & 1,743 & 26.1 & 4.82 & 7.4 \\
            \midrule
            \multirow{2}{*}{Full} 
            & Plosives     & 40 & 1,643 & 25.9 & 4.81 & 7.1 \\
            & Front vowels & 40 & 1,686 & 26.3 & 4.79 & 7.1 \\
            \midrule
            \multirow{2}{*}{Limited} 
            & Plosives     & 34 & 865 & 25.1 & 0 & 7.2 \\
            & Front vowels & 35 & 875 & 22.4 & 0 & 7.1 \\            
            \bottomrule
        \end{tabular}
    }
    \vspace{-10pt} %
\end{table}

\section{Experiments}
\label{section:experiments}
This study focuses on two questions: whether a model can acquire target phonemes unseen in pre-training through fine-tuning, and how pre-trained knowledge benefits this process. 
We conducted two experiments under simulated phoneme-controlled settings and real-speech cross-lingual transfer settings.
In the simulated phoneme-controlled setting (Section~\ref{subsection:effectiveness_pretraining}), we first verified the effectiveness of pre-trained phoneme knowledge for phoneme addition using phoneme-controlled synthetic corpora that kept the language and speaker fixed.
This simulation can reduce confounding factors such as language, speaker, and data domain differences, which often arise in cross-lingual experiments. 
In the real-speech cross-lingual transfer setting (Section~\ref{subsec:cross_lingual}), we further validated whether our findings generalize to a cross-lingual transfer learning setting with real speech datasets.

To assess the effect of training dataset size, both experiments compared the performance of fine-tuned models with scratch-trained models using various amounts of training data ($100$, $300$, $500$, $800$, $1,000$, and $2,000$ utterances, each randomly selected from the respective target datasets).

\subsection{Experimental settings}
\subsubsection{Model Training Configuration}
We used the implementation of Conformer-FastSpeech2 (CFS2)~\cite{cfs2, conformer, fastspeech2} and its configurations from ESPnet~\cite{espnet}. 
For the pre-training, we followed the training configuration from ESPnet\footnote{\url{https://github.com/espnet/espnet/blob/master/egs2/vctk/tts1/conf/tuning/train_xvector_conformer_fastspeech2.yaml}}, except for using speaker ID conditioning instead of the x-vector conditioning.
Following the baseline of previous studies~\cite{chen2019end, do2022text, do2023strategies}, we adopted the naive fine-tuning approach: the phoneme inventory is expanded to include unseen target phonemes, with only their embeddings randomly initialized.
For fine-tuning, we removed speaker conditioning to train the model as a single-speaker model and reduced the learning rate to one-tenth of its original value. 
Through the preliminary experiments, we confirmed that varying the learning rate had a limited effect on the fine-tuned model performance.

As a baseline, we also trained a single-speaker CFS2 model from scratch (random initialization) on the same dataset as fine-tuning, denoted as ``Scratch.''
For waveform generation, we used a pre-trained HiFi-GAN~\cite{hifigan} vocoder from the ParallelWaveGAN repository\footnote{\url{https://github.com/kan-bayashi/ParallelWaveGAN}} trained on the VCTK dataset.

\subsubsection{Evaluation metrics}
To assess the effectiveness of the pre-trained knowledge from various perspectives, we evaluate phoneme addition performance and naturalness.
For phoneme-addition evaluation, we use a wav2vec 2.0-based phoneme recognition model~\cite{w2v2phoneme_recognizer} to compute the target phoneme error rate (Target PER).
Inspired by the biased-WER~\cite{le21_interspeech_context_bias, huang23d_interspeech_b_wer, b_wer}, Target PER measures how accurately the model generates newly added phonemes by calculating PER solely on the target phonemes.
For naturalness evaluation, we employed the pre-trained UTMOS model\footnote{\url{https://github.com/sarulab-speech/UTMOS22}}~\cite{utmos}, which simulates the mean opinion score (MOS) of synthesized speech.

\subsection{Simulated Phoneme-Controlled Settings}
\label{subsection:effectiveness_pretraining}
This experiment investigates how pre-trained knowledge contributes to phoneme addition during fine-tuning using simulated phoneme-controlled datasets.
Since all utterances were generated by the same synthesis pipeline, acoustic conditions were consistent across corpora.
This allows a focused investigation of phoneme addition across various conditions, including training dataset sizes and target phoneme types.

\subsubsection{Prepared TTS model for simulated dataset creation}
\label{par:tts_model_for_corpus_generation}
To build the synthetic speech--phoneme paired datasets, we used a prepared TTS model trained on the VCTK dataset, which contains approximately 44 hours of English speech from 108 speakers~\cite{vctk}.
For its training, we converted text from the VCTK dataset to IPA-style phoneme sequences using espeak-ng.
Using this model, we created speech--phoneme paired datasets by synthesizing audio for the LLM-generated ``Limited'' and ``Full'' text corpora, as described in Section~\ref{sec:method}.

\subsubsection{Simulated datasets with phoneme-controlled speech}
\label{par:synthetic_dataset}
We created speech--phoneme paired datasets by synthesizing audio for the LLM-generated ``Limited'' and ``Full'' text corpora using the prepared TTS model described in Section~\ref{par:tts_model_for_corpus_generation}:

\noindent\textbf{Limited dataset} (for pre-training): Contains 107,000 utterances (55 hours) from 107 speakers excluding the target speaker (a female native English speaker, p299), which is a larger scale than the VCTK dataset.
This dataset excludes specific target phonemes intended to be added during fine-tuning.

\noindent\textbf{Full dataset} (for fine-tuning): Contains 2,000 utterances from the target speaker (p299).
This dataset contains all English phonemes, including the target phonemes.

We first created a pre-trained TTS model trained on the Limited dataset, and subsequently fine-tuned this model with the Full dataset.
To investigate how pre-trained knowledge contributes to phoneme addition during fine-tuning, we compared fine-tuning against training a model from scratch on the Full dataset.

\subsubsection{Target Phonemes and Evaluation}
To evaluate the influence of target phoneme type on phoneme addition performance, we conducted experiments with two different groups of target phonemes:
\textbf{Plosive Consonant} (\textipa{/p/, /b/, /t/, /d/, /k/, and /g/}) and \textbf{Front vowels} (\textipa{/i/, /I/, /eI/, /E/, and /\ae/}).
We evaluated synthesized speech using a VCTK text subset of 1,000 sentences, each containing at least one target phoneme. 
These sentences were not included in the training dataset for pre-training and fine-tuning.
For PER evaluation in English, we used a wav2vec 2.0-based phoneme recognition model fine-tuned on VCTK dataset.

\subsection{Real-speech Cross-lingual Transfer Setting}
\label{subsec:cross_lingual}

In this Section, we further validated whether our findings generalize to a practical cross-lingual transfer learning scenario.
Specifically, we conducted a cross-lingual phoneme addition experiment from English to Japanese using real speech datasets.

\subsubsection{Real-speech English and Japanese Speech Datasets}
As the source-language (English) dataset for pre-training, we used the VCTK dataset.
As the target-language (Japanese) corpus, we used the single-speaker JSUT dataset~\cite{sonobe2017jsut}.

\noindent\textbf{VCTK (English) dataset} (for pre-training): Contains approximately 44 hours of English speech from 108 native speakers.

\noindent\textbf{JSUT (Japanese) dataset} (for fine-tuning): Contains approximately 10 hours of Japanese speech from a single native female speaker. This experiment used the basic5000 set of the JSUT dataset.
For Japanese text preprocessing, we converted Japanese transcriptions to phoneme sequences using pyopenjtalk~\footnote{\url{https://github.com/r9y9/pyopenjtalk}}and mapped the outputs to IPA symbols.

We created a pre-trained English TTS model trained with the VCTK dataset, and subsequently fine-tuned this model with the randomly sampled subset of the JSUT dataset. 
To investigate how English pre-trained knowledge contributes to Japanese phoneme addition during fine-tuning, we compared the Japanese fine-tuned model from an English pre-trained model against a Japanese scratch-trained model.

\subsubsection{Target Phonemes and Evaluation}
For cross-lingual phoneme addition, during fine-tuning, we expanded the base English phoneme vocabulary of 40 symbols by adding 20 \textbf{Japanese-specific phonemes} unseen during English pre-training. Specifically, the added phonemes are grouped as follows: vowels (\textipa{/a/, /e/, /o/, /\textturnm/}), stops (\textipa{/b\super j/, /c/, /d\super j/, /\textbardotlessj/, /p\super j/}), affricates (\textipa{/ts/, /t\textctc/, /d\textctz/}), fricatives (\textipa{/\c{c}/, /\textctc/, /\textphi/}), nasals (\textipa{/m\super j/, /\textltailn/, /\textscn/}), flap (\textipa{/\textfishhookr\super j/}), and geminate consonant (/Q/).
This experiment regarded these 20 Japanese-specific phonemes as target phonemes.
For evaluation, we used 100 Japanese sentences from the JVS corpus~\cite{takamichi2019jvs} as test texts, ensuring that they were not used for training. 
For PER evaluation in Japanese, we used a wav2vec~2.0-based phoneme recognition model fine-tuned on JSUT basic5000 dataset.

\begin{figure}[t]
   \centering
    \begin{minipage}[b]{1.0\linewidth}
      \centering
      \centerline{\includegraphics[width=\hsize]{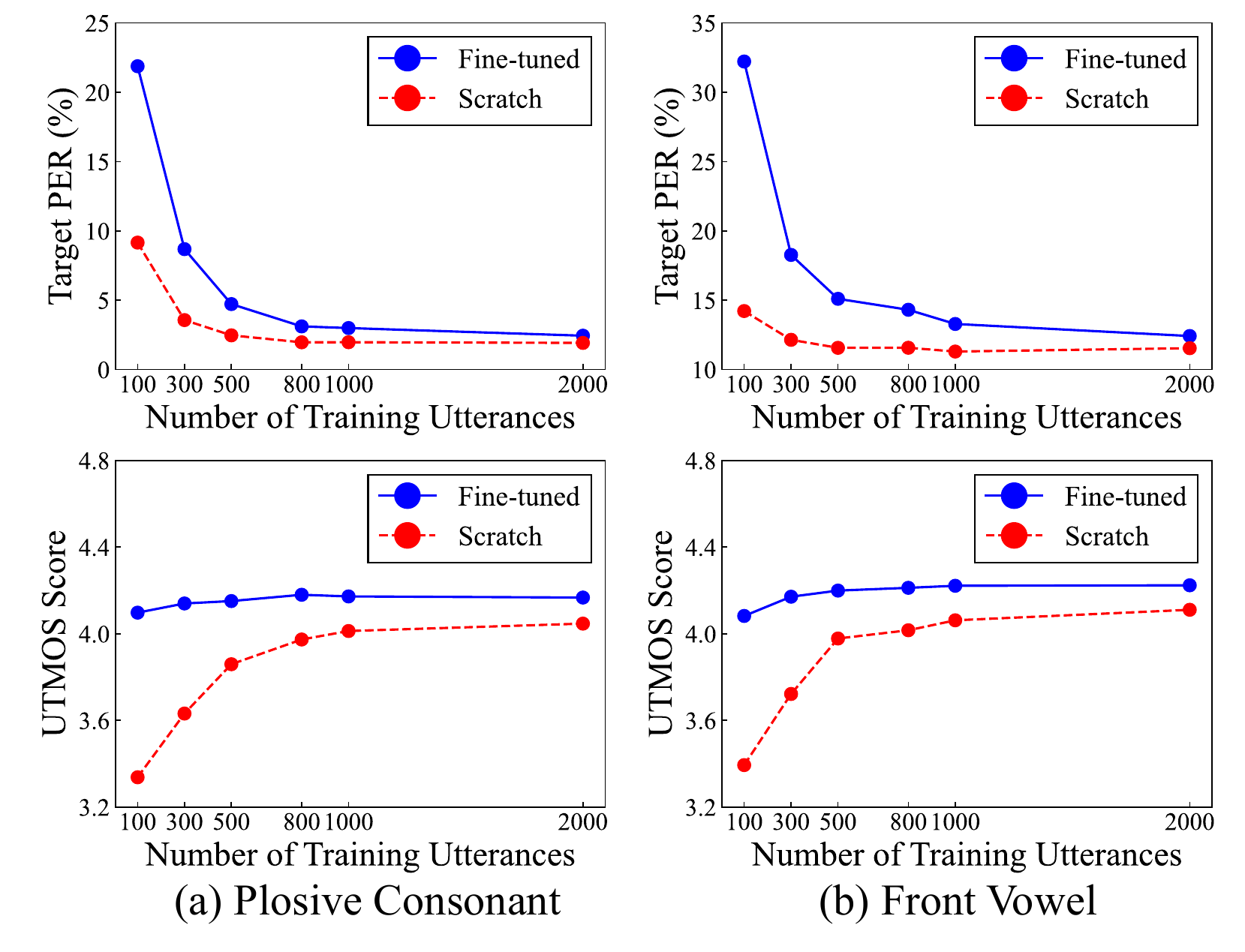}}
    \end{minipage}
        \vspace{-15pt} %
\caption{Target PER (\%) (top) and UTMOS score (bottom) comparison between fine-tuning (blue solid) and scratch training (red dashed) for (a) plosive consonants and (b) front vowels across various dataset sizes.}
    \vspace{-10pt} %
    \label{fig:plosive_vs_frontvowel}
  \end{figure}

\section{Experimental Results}
\label{sec:experimental_results}

\subsection{Results in Simulated Phoneme-Controlled Setting}
\label{subsection:result_effectiveness_pretraining}
Figure~\ref{fig:plosive_vs_frontvowel} shows Target PER and UTMOS score comparisons between fine-tuning and training from scratch for plosive consonants and front vowels across various dataset sizes (100, 300, 500, 800, 1,000, and 2,000).
Across both target phoneme types, training from scratch achieved comparable or better PER than fine-tuning in both cases.
In other words, while scratch training was required to learn the ability to generate speech for all phonemes from scratch, fine-tuning required approximately the same or more amount of training data as scratch training to achieve comparable PER, regardless of the target phoneme type.
From the perspective of phoneme accuracy, this result suggests that the pre-trained phoneme knowledge is not effectively leveraged during fine-tuning phoneme addition.
This finding is contrary to the common expectation that the linguistic and acoustic knowledge acquired during pre-training mainly contributes to the fine-tuning task.
In contrast to the Target PER results, we should note that the pre-trained knowledge contributes to the perceptual quality of the synthesized speech.
Across both target phoneme types, fine-tuning consistently achieved better naturalness than scratch training across all data sizes.
These results indicate that while the pre-trained phoneme knowledge does not enhance new phoneme addition, it contributes to improving the naturalness of the synthesized speech.

Furthermore, we analyzed spectrograms of synthesized speech by both the fine-tuned model and a model trained from scratch after phoneme addition of the plosive consonants.
In low-resource conditions (100 utterances), scratch training could produce plosive closure patterns more clearly, while both approaches became comparable in sufficient data conditions (more than 1,000 utterances).
Notably, the scratch-trained model could generate the target phonemes as well as other phonemes. In contrast, while the fine-tuned model could already produce seen phonemes accurately, it struggled to generate the new target phonemes.
These observations suggest the differences in training constraints. 
Specifically, scratch training can learn all phonemes jointly without any constraints.
In contrast, fine-tuning acquires the target phonemes while preserving previously learned phoneme knowledge, which makes it harder to acquire newly added phonemes.

\subsection{Results in Real-speech Cross-lingual Transfer Setting}
\label{subsection:result_cross_lingual}

\begin{figure}[t]
\captionsetup{font={footnotesize}}
  \begin{minipage}[b]{1.0\linewidth}
    \centering
    \includegraphics[width=\hsize]{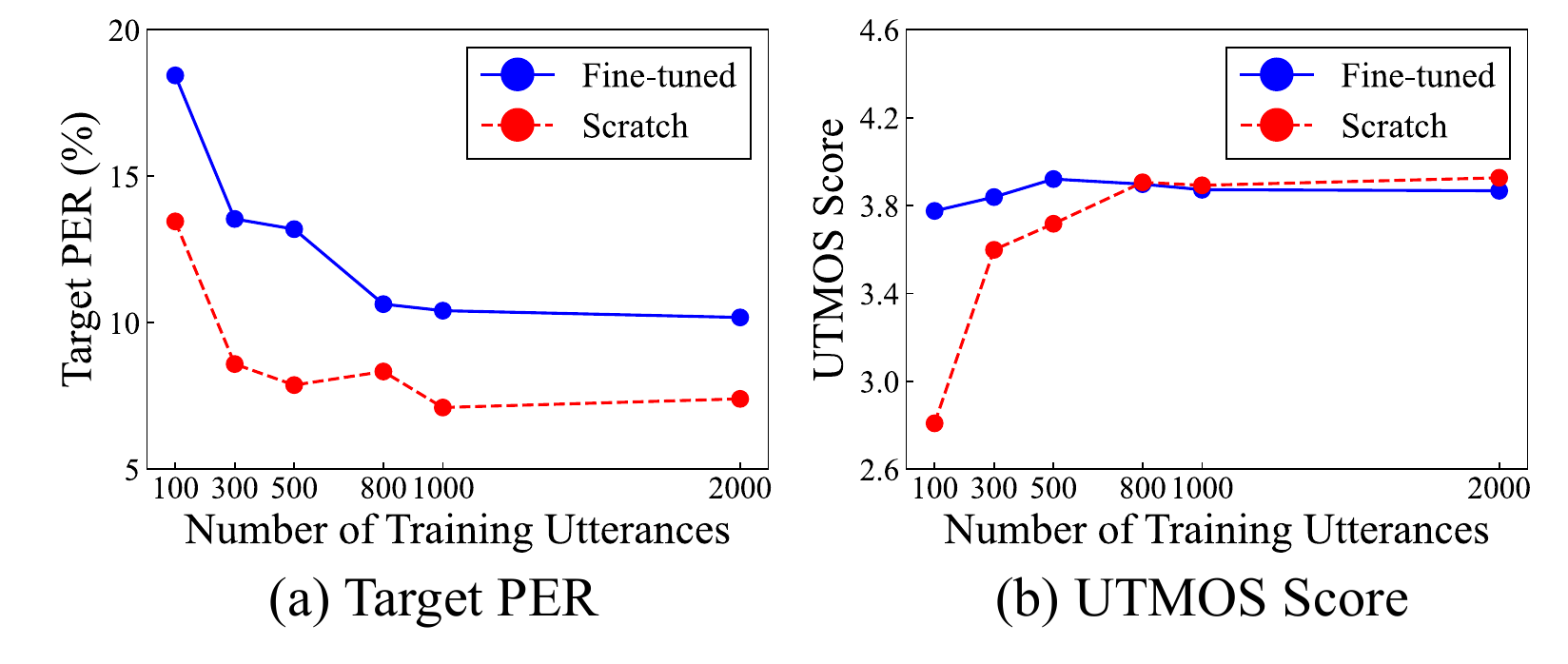}
  \end{minipage}
  \vspace{-15pt}
  \caption{(a) Target PER (\%) on Japanese-specific target phonemes and (b) UTMOS score in a real-speech cross-lingual transfer learning setting (English VCTK $\rightarrow$ Japanese JSUT), comparing fine-tuning (blue solid) and scratch training (red dashed) across various dataset sizes.}
  \label{fig:vctk_jsut}
  \vspace{-10pt} %
\end{figure}

Figure~\ref{fig:vctk_jsut} shows the results in the real-speech cross-lingual experiment from English (VCTK) to Japanese (JSUT).
For Target PER on Japanese-specific phonemes, Figure~\ref{fig:vctk_jsut} (a) shows that training from scratch consistently achieved better Target PER than fine-tuning across all data sizes.
This trend is consistent with the observations in the controlled simulation experiment described in Section~\ref{subsection:result_effectiveness_pretraining}.
This result suggests that, even in a real-speech cross-lingual transfer learning scenario, the pre-trained ability to generate seen phonemes offers limited benefit for acquiring unseen target phonemes.

In contrast, the naturalness evaluations in Figure~\ref{fig:vctk_jsut} (b) show an advantage of pre-training in a low-resource fine-tuning scenario.
Fine-tuning achieved higher or comparable UTMOS scores than scratch training, especially in lower-resource conditions (100, 300, and 500 utterances).
These results indicate that pre-training mainly contributes to naturalness improvement, whereas it offers limited benefit for acquiring unseen new phonemes compared to scratch training.
  
\section{Conclusions}
\label{sec:conclusions}

This study investigated how pre-trained phoneme knowledge contributes to the phoneme addition process through fine-tuning.
We conducted experiments under two settings: (1) a phoneme-controlled simulation using LLM-generated corpora to enable the comprehensive evaluation without considering confounding factors, and (2) a real-speech cross-lingual transfer scenario (English to Japanese) to confirm the validity of our simulation findings.
Across both settings, experimental results showed that TTS models successfully acquired new phonemes through fine-tuning regardless of target phoneme types.
However, while fine-tuned models achieved better UTMOS scores, they only achieved comparable or worse PER compared with the models trained from scratch.
Contrary to the common expectation, these results indicate that pre-trained phoneme knowledge does not always benefit phoneme addition.
While the contribution of pre-trained phoneme knowledge is limited for phoneme addition, pre-training mainly benefits speech naturalness improvement. 
To leverage pre-trained phoneme knowledge more effectively, we can explore several promising directions.
One approach is using pre-trained models with broader phoneme inventories to cover more target phonemes, while another involves an auxiliary loss designed to encourage learning new phonemes.
Future work will explore developing methods for leveraging pre-trained phoneme knowledge more effectively to enhance phoneme addition even with tiny datasets.

\vfill\pagebreak
\section{Acknowledgments}
This work was supported in part by JSPS KAKENHI Grant Number 26H02530, and in part by the BRIDGE Program (R7-H05), implemented by the Cabinet Office, Government of Japan.

\section{Generative AI Use Disclosure}
We used ChatGPT 5.2 Pro and Claude Opus 4.6 for polishing manuscripts.
\bibliographystyle{IEEEtran}
\bibliography{mybib}

\end{document}